\documentclass[prl,twocolumn,superscriptaddress,showpacs]{revtex4}
\usepackage{amsmath}
\usepackage{bm}
\usepackage{epsfig}
\usepackage{color}
\usepackage{float}
\usepackage{bm}
\usepackage{graphicx}
\newcommand{\CC}{\mathcal{C}}
\newcommand{\eps}{\epsilon}

\begin{document}

\title{Evidence for a disordered critical point in a glass-forming liquid}

\author{Ludovic Berthier}

\affiliation{Laboratoire Charles Coulomb, 
UMR 5221 CNRS-Universit\'e de Montpellier, Montpellier, France}

\author{Robert L. Jack}
\affiliation{Department of Physics, University of Bath, 
Bath, BA2 7AY, United Kingdom}

\newcommand{\rlj}[1]{{\color{blue}#1}}

\date{\today}


%05.10.-a 	Computational methods in statistical physics and nonlinear dynamics (see also 02.70.-c in mathematical methods in physics)
%05.20.Jj 	Statistical mechanics of classical fluids (see also 47.10.-g General theory in fluid dynamics)
%64.70.Q- 	Theory and modeling of the glass transition 

\begin{abstract}
Using computer simulations of an atomistic glass-forming liquid, 
we investigate the fluctuations
of the overlap between a fluid configuration and a quenched 
reference system. We find that
large fluctuations of the overlap develop as temperature decreases, 
consistent with the existence
of the random critical point that is predicted by effective field theories. 
We discuss the scaling
of fluctuations near the presumed critical point, comparing the observed 
behaviour with that of
the random-field Ising model. We argue that this critical point directly 
reveals the existence of an
interfacial tension between amorphous metastable states, a quantity 
relevant both for equilibrium
relaxation and for nonequilibrium melting of stable glass configurations.
\end{abstract} 

\maketitle

In the search for a theory of the glass transition, an important 
question is whether the slow dynamics of supercooled liquids 
can be explained in terms 
of the temperature evolution of `universal' thermodynamic quantities, 
independent on the details of the material~\cite{Berthier_Biroli_2011}. 
Two-body density correlators, which are central to 
the theory of simple liquids~\cite{hansen_theory_1986}, fail for viscous 
liquids \cite{gilles}. In the framework of the 
random first-order transition (RFOT) theory \cite{rfot_review}, 
slow dynamics is described 
in terms of infrequent transitions between amorphous free energy minima, 
separated by large barriers. Structural relaxation is interpreted as the
nucleation of one metastable state into another \cite{thirumalai__1989}. 
This physical picture is supported 
by mean-field calculations of a free-energy $V(Q)$.
Here, the overlap $Q$ measures the similarity
between pairs of configurations: it acts as an order parameter for the glass
transition \cite{franz_phase_1997}. 
This `Landau free energy' describes the overlap fluctuations 
and links statics to dynamics. 
Because they embody high-order
density correlations, overlap fluctuations are key candidates 
to construct a thermodynamic theory of the glass transition and, 
as such, are currently the focus of a large interest 
\cite{cammarota_phase-separation_2010,moore2012,berthier_overlap_2013,sconf,long_giulio,FP2013,tarzia}. 
The central challenge, tackled here, 
is to understand whether the rigorous results obtained in the 
mean-field limit are relevant for realistic, finite dimensional liquids. 
 
At mean-field level, $V(Q)$ allows for a compact description of the
liquid-glass phase transition occurring
at the Kauzmann temperature, $T_K$, for which $Q$ is the relevant 
order parameter.  
By introducing additional external fields, such as a coupling 
$\epsilon$ to a reference copy of the system, the glass transition 
at $(T=T_K, \epsilon=0)$ transforms 
in a first-order transition line ending at a critical point at
$(T_c, \epsilon_c > 0)$ \cite{kurchan93,franz_phase_1997,long_giulio}. 
Because $T_c > T_K$, this critical point might be 
more easily accessible than $T_K$. This scenario
occurs in some disordered spin models but has not been demonstrated in 
equilibrium calculations on atomistic systems (see 
\cite{franz_parisi1998,cardenas_constrained_1999,cammarota_phase-separation_2010} for earlier work).  
Here, we present free-energy calculations that 
provide direct evidence for such a critical point in a realistic 
glass-former, and find a critical behaviour consistent with 
the universality class of the random field ising model (RFIM), 
in agreement with theory \cite{FP2013,tarzia}. Further motivation 
to analyse this critical point is that its absence would directly 
establish that a Kauzmann transition does not occur 
at any $T_K>0$.
(The reverse is not true: a finite $T_c$ does not imply 
a finite $T_K$.) The existence of a first-order transition 
line at $T<T_c$ is even more significant for 
understanding glassy phenomenology because it is associated with phase 
coexistence between equivalent metastable states. In particular, 
the corresponding interfacial tension, $\Upsilon(T)$, is an important piece
of the RFOT theory \cite{thirumalai__1989,Bouchaud-JCP2004}, 
but even establishing its existence  
has proven difficult in computational studies \cite{cammarota_phase-separation_2010,chiara1,chiara2}. 
Our approach bypasses these problems and paves the way for 
direct quantitative determinations of $\Upsilon$.
As a first step, we analyse cases where stable glass 
states coexist with liquid states, with a large barrier 
between them that we attribute to an interfacial free energy. 
This finding is relevant to the current effort to understand
the melting process of ultrastable glasses 
\cite{inactive2011,sepulveda2014,hocky2014}. 

We consider a well-studied binary mixture 
of Lennard-Jones particles \cite{kob_testing_1995}, 
which exhibits glassy dynamics below $T \approx 1.0$. The unit of length is
the diameter of the larger particles, the unit 
of energy is the pair-interaction scale, the Boltzmann constant is unity, 
and the total number of particles is $N$.  
The position of particle $i$ is $\bm{r}_i$, and $\CC=(\bm{r}_1,
\bm{r}_2,\dots,\bm{r}_N)$ denotes a configuration of the system.  
Following \cite{franz_phase_1997}, 
we simulate two copies of the system.
First, a \emph{reference configuration} $\CC_0$ is drawn at
random from the equilibrium state of the model at temperature $T'$.  
Then, we perform Monte-Carlo (MC) simulations on a second
configuration $\CC$, which is biased by a field $\eps$ to lie close to 
$\CC_0$. Specifically, our MC method samples configurations $\CC$ according to 
the Boltzmann weight for the Hamiltonian 
$E(\CC) - \eps Q(\CC,\CC_0)$, where $E$ is the potential energy and 
$Q(\CC,\CC') = \frac{1}{N} \sum_{ij} \Theta( a - |\bm{r}_i - \bm{r}_j'| )$
is the overlap between configurations $\CC$ and $\CC'$.  Here, 
$\Theta(x)$ is the Heaviside function and we take $a=0.3$.
Efficiently sampling this Hamiltonian is challenging, and 
we adopt the strategy detailed in \cite{berthier_overlap_2013,sconf}
which combines parallel tempering with umbrella sampling and 
reweighting techniques. Using this approach we were able to study 
system sizes $N=150$, $256$ for temperatures $T \geq 0.55$. 
  
The temperatures $T$ and $T'$ may be different 
(${\cal C}_0$ need not even be thermalized), 
but the case $T=T'$ has special importance.
For $\eps=0$, $\CC$ and $\CC_0$ are independent. 
As $\eps$ increases, the configurations become more 
similar and their mutual overlap grows. Our goal 
is to investigate the presence of singularities in overlap
fluctuations in the $(\epsilon,T)$ plane.  
To this end, a central role is played by the 
($\eps$-dependent) distribution of the overlap, $P(Q)$. This distribution 
is obtained by first performing a thermal average over $\CC$, 
and then averaging the results over reference configurations $\CC_0$ (we
typically use 60 independent configurations for each  
$T$ and $N$). This double average is indicated 
with simple brackets $\langle\cdot\rangle$. 

We emphasize that the reference 
configuration $\CC_0$ is fixed (or `quenched') within each simulation, 
and sampling is restricted to configuration $\CC$.
This approach is relevant for glassy systems, 
because we expect $\CC_0$ to be representative of equilibrium
metastable states at temperature $T'$.  
The configuration $\CC$, at temperature $T$, 
may then either occupy the same metastable 
state (when $Q$ is large) or a different one (when 
$Q$ is low). If phase coexistence occurs at $T'=T$, 
then {\it it takes place between equivalent metastable states}.
The `annealed' case in which both configurations fluctuate 
is qualitatively different because the high-$Q$ configurations 
are not representative equilibrium states
\cite{monasson,berthier_overlap_2013,mezard99,parisi_liquid-glass_2014,garrahan2014,bomont_probing_2014}.

\begin{figure}
\psfig{file=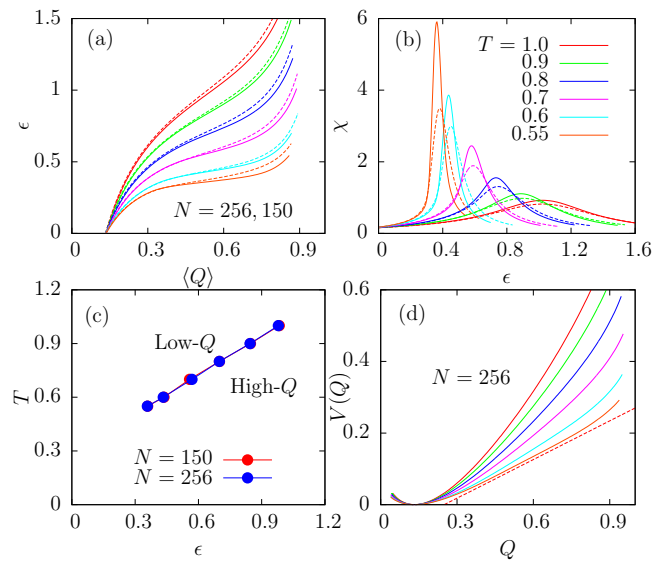,width=8.5cm}
\caption{(a) Dependence of the average overlap $\langle Q \rangle$ on the
field $\eps$ along various isotherms for $N=256$ (full) and $N=150$ 
(dashed). The behaviour is reminiscent of 
isotherms in a liquid-vapor system. 
(b) Total susceptibility $\chi(\epsilon,T)$ for the same parameters
showing increasing fluctuations as temperature is reduced 
and system size is increased.  
(c) Temperature-dependence of the field $\eps^*$ 
that maximises $\chi$, indicating the boundary between low-$Q$ and 
high-$Q$ regimes.  
(d) Free energy $V(Q)$ for the same temperatures as in (a, b) 
for $N=256$, with a dashed line indicating linear behavior.}
\label{fig1}
\end{figure}

In Fig.~\ref{fig1} we show results for $T' = T$.
The isotherms $\langle Q \rangle (\epsilon,T)$ shown in
Fig.~\ref{fig1}(a) recall the relationship between pressure
and density in a liquid-vapor system. 
On lowering the temperature, they become increasingly flat. 
This evolution is more pronounced for the larger system.
Below the critical point, one would observe, in the thermodynamic limit,
horizontal tie-lines associated with phase coexistence.
 Fig.~\ref{fig1}(b) shows the 
susceptibility $\chi (\epsilon,T) = N \left( \langle Q^2 \rangle 
- \langle Q \rangle^2 \right)$, which exhibits a pronounced maximum at a 
field $\epsilon^*(T)$ that shifts to lower values when $T$ decreases,
suggesting a reduction in the thermodynamic cost 
of localizing the system near a reference configuration (related to the configurational entropy).
Interestingly $\chi(\epsilon^*,T)$ grows rapidly as $T$ is lowered,
and also increases  strongly with system size. 
This behavior is expected if the system approaches 
a critical point, but can also be obtained if a correlation length
larger than the system size develops. 
The evolution of  $\eps^*(T)$ is shown in   
Fig.~\ref{fig1}(c) in the $(\eps,T)$ phase diagram. 
In the liquid-vapor analogy, this corresponds to the critical isochore
and, for temperatures above the critical point, to a Widom 
line \cite{domb1972}. The observed linear decrease of $\epsilon^*$ 
with $T$ is consistent with mean-field theories.
 
Finally, Fig.~\ref{fig1}(d) shows the free energy 
$V(Q) = (-T/N) \log P(Q)$ measured at various $T$, 
for $\epsilon = 0$. It has the usual convex form at
high $T$ with a minimum near $Q \approx 0$. For lower $T$ 
and intermediate values of $Q$, 
its curvature decreases markedly, 
consistent with the increasing variance of $Q$. 
For $T<T_c$, one expects $V(Q)$ to acquire
a region of zero curvature, due to Maxwell's construction, 
similar to the behavior shown at the lowest studied temperature $T=0.55$. 
Whereas the value of $V(Q)$ in the high-overlap regime provides 
a direct estimate of the configurational entropy \cite{sconf}, 
the postulated surface tension $\Upsilon(T)$ between states 
cannot be obtained directly from $V(Q)$.

\begin{figure}
\psfig{file=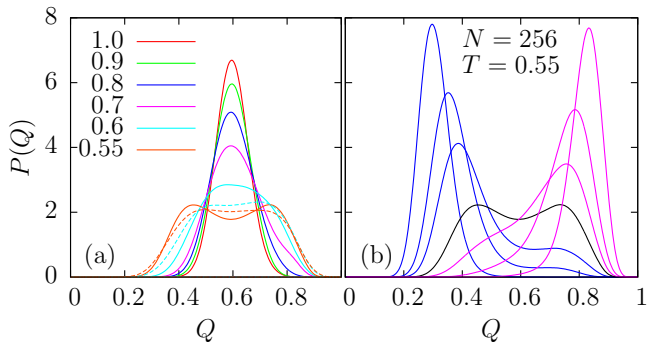,width=8.5cm}
\caption{(a) Overlap distributions $P(Q)$ measured along 
the $\eps^*(T)$ line for $N=256$ (full lines); dotted lines are 
for $N=150$ and $T=0.6$ and 0.55. For the lowest 
temperature, a bimodal structure appears.
(b) Behaviour of $P(Q)$ for various $\eps$, increasing from 
low-$\epsilon$ (blue) to $\eps=\eps^*$ (black) and large-$\epsilon$ (purple)
for $N=256$ and $T=0.55$.}
\label{fig2}
\end{figure}

To gain further insight, we analyse the 
overlap distribution, $P(Q)$, because it can 
better reveal critical fluctuations and phase coexistence. 
We show in Fig.~\ref{fig2}(a) the temperature evolution 
of these distributions evaluated along the $\eps^*(T)$ line.
The distribution is narrow and essentially Gaussian 
at high temperature, $T \geq 0.7$. It becomes 
broader, develops a non-Gaussian flat center, and 
eventually becomes weakly bimodal for $T=0.55$, $N=256$. 
Bimodality is more pronounced for the largest system. 
To investigate this $N$-dependence, 
we calculated the Binder cumulant ratio
$B=\langle ( Q - \langle Q\rangle)^2\rangle^2 / \langle ( Q - \langle 
Q\rangle)^4\rangle$. On a first-order transition line,
$B$ increases with $N$, tending to unity as $N \to \infty$ whereas 
above $T_c$ it decreases with $N$ towards $\frac13$.  
At $T_c$, $B$ is size-independent with a limiting 
value intermediate between $\frac13$ and unity.
At high temperatures, we find $B \approx \frac13$
as expected, and for $T=0.55$ and $N=(150,256)$, we get $B=(0.48,0.52)$, 
consistent with this temperature being close to $T_c$, although our
range of system sizes is rather small.
For $T=0.55$, $N=256$ we show $P(Q)$ 
in Fig.~\ref{fig2}(b), which evolves from a unimodal
distribution at low-$Q$ when $\epsilon < \epsilon^*$
to a similarly unimodal distribution at high $Q$ when $\epsilon > \epsilon^*$,
with an intermediate bimodal distribution at $\epsilon^*$.
This evolution is qualitatively consistent with the crossing 
of a first-order-like transition line in a finite-size system, but 
systematic finite-size scaling is needed to draw conclusions 
about the thermodynamic limit.

\begin{figure}
\psfig{file=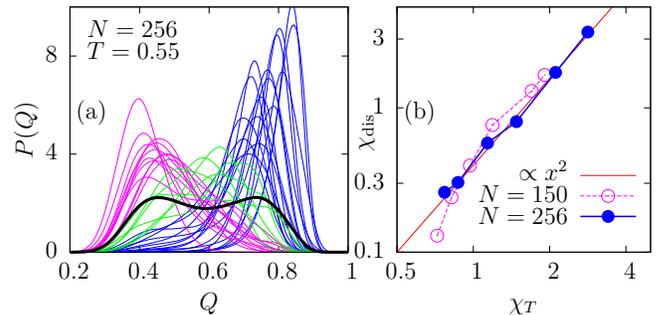,width=8.5cm}
\caption{(a) Sample-to-sample fluctuations 
of the distributions $P(Q)$ for $T=0.55$, $N=256$
evaluated at the average $\epsilon^*$.
Colors indicate whether the samples are predominately in 
high-, intermediate, or low-$Q$ states. 
As expected for RFIM behaviour, most distributions are 
unimodal, averaging to a total bimodal distribution (black).
(b) Temperature dependence of the susceptibilities $\chi_{\rm dis}$ 
and $\chi_T$ measured at $\eps^*$. The solid
line is the RFIM prediction, Eq.~(\ref{equ:chichi}).}
\label{fig4}
\end{figure}

Random critical points in systems with quenched 
disorder have distinct properties from standard liquid-vapor 
transitions \cite{Young_1998}. In particular, the system may behave differently
for different realizations of the disorder \cite{Fisher-PRL-1986}, 
in our case different reference configurations $\CC_0$.
It is thus useful to decompose averages, using 
$\langle \cdot \rangle_{\CC_0}$ for a
thermal average over $\CC$ at fixed $\CC_0$, 
and an overbar $\overline{(\cdot)}$ for a disorder average over $\CC_0$. 
We find that the isotherms $\langle Q \rangle_{\CC_0}$
have strong sample-to-sample fluctuations. The overlap grows more or 
less abruptly with $\epsilon$ for different samples,
so that the critical field $\epsilon^*_{\CC_0}$ is a distributed quantity. 
In Fig.~\ref{fig4}(a) we show that at low temperature, $T=0.55$,
the distribution of $\eps^*_{\CC_0}$ is so broad 
that the distributions $P(Q)$ for most samples 
at the average value $\epsilon^* = \overline{\eps^*_{\CC_0}}$ are dominated by 
either their small-$Q$ or large-$Q$ peaks. 
Most of these distributions do become 
bimodal when the field is ajusted to $\epsilon^*_{\CC_0}$ 
for each sample separately. These results strongly suggest that the 
quenched disorder plays a significant role in the overlap fluctuations.

In mapping the supercooled liquid to the RFIM,
the configuration $\CC$ corresponds to an Ising model in a random 
magnetic field, and the specific realisation of 
the disorder corresponds to 
$\CC_0$. In the RFIM, each realisation of the random field slightly 
favors either positive or negative magnetisation. As a 
result, each sample typically has only one thermally-populated 
state \cite{Fisher-PRL-1986}, in constrast to the pure case 
for which both states are equally likely. In large systems, 
this disorder effect dominates the critical fluctuations, 
which changes the universality class from Ising to RFIM.

For quantitative analysis, we decompose fluctuations
into thermal, $\chi_T=N\overline{(\langle 
Q^2\rangle_{\CC_0} -\langle Q \rangle_{\CC_0}^2)}$, 
and disorder parts, $\chi_{\rm dis}=N(\overline{\langle Q\rangle_{\CC_0}^2}-
\overline{\langle Q\rangle_{\CC_0}}^2)$.
Clearly, $\chi = \chi_{\rm dis} + \chi_T$. 
Numerical results, shown in Fig.~\ref{fig4}(b), indicate that 
for the largest system the disorder fluctuations increase 
much more rapidly than thermal ones, 
\begin{equation}
{\chi_{\rm dis} \propto \chi_T^2,}
\label{equ:chichi}
\end{equation}
implying that the disorder should eventually completely dominate the 
behaviour of the total susceptibility, $\chi \approx \chi_{\rm dis} 
\gg \chi_T$. Physically, Eq.~(\ref{equ:chichi}) can be 
understood by assuming that fluctuations are completely
dominated by the sample fluctuations of the field $\eps^*_{\CC_0}$.
In that case, standard results for ensemble-dependence of 
fluctuations \cite{schwartz85,science05,berthier_spontaneous_2007}
yield $\chi_{\rm dis} \approx N\, \mathrm{Var}(\eps^*) 
\chi_T^2/T^2$, where $\mathrm{Var}(\eps^*)$
is the variance of $\eps^*_{\CC_0}$ among samples.
The resulting scaling relation (\ref{equ:chichi})
is characteristic of the RFIM universality class. 
It is exact in mean-field analysis \cite{FP2013}, and  
the deviations predicted from nonperturbative renormalization
treatments in $d=3$ \cite{gillesRG} are too small to be 
numerically observable \cite{fytas}.
Interestingly, the ratio $\chi_{\rm dis}/\chi_T^2$ provides 
a measure of the (effective) variance of the field,
which is a dimensionless measure of the strength of the disorder in the 
effective RFIM description of the critical point \cite{tarzia}. 
It would be interesting 
to perform similar measurements for other models to perform
a quantitative comparison of effective theories 
for different liquids \cite{wolynesRFIM}.

\begin{figure}
\psfig{file=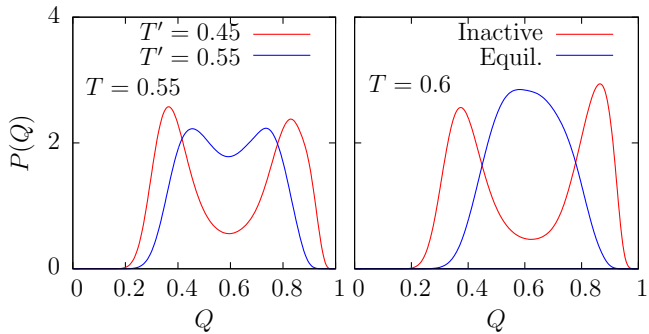,width=8.5cm}
\caption{(a) Comparison of distributions $P(Q)$ and $T=0.55$, for 
$T'=T$ and $T'=0.45$, at their corresponding fields $\eps^*$.
(b) Comparison at $T=0.6$ between equilibrated $\CC_0$ 
at $T'=0.6$ and (biased) inactive $\CC_0$.
In both cases, the more stable states are associated
with strongly bimodal $P(Q)$, indicating that coexistence between these 
low-energy states and the equilibrium fluid incurs a significant 
interfacial free energy cost. }
\label{fig3} 
\end{figure}

The distribution $P(Q)$ allows an
estimation of the surface tension $\Upsilon(T)$, provided the 
first-order transition regime characterized by well-separated peaks 
can be accessed. Unfortunately, the computational effort required to sample 
$P(Q)$ for $T \ll 0.55$ is prohibitive.
To circumvent this difficulty,
we exploit the flexibility offered by the independent 
sampling of configurations $\CC$ and $\CC_0$. 
When $T' \neq T$, mean-field theory predicts that 
the three-parameter space $(\eps,T,T')$ contains
a line of {RFIM} 
critical points.
The phase diagram in Fig.~\ref{fig1}(c) corresponds to the 
$T'=T$ plane in that space, but fixing for instance $T'=0.45$
should yield another critical point $(\tilde{\eps}_c,\tilde{T}_c)$.  
If $\tilde{T}_c > T_c$, then
sampling temperatures below $\tilde{T}_c$ should be easier. 
This is demonstrated in Fig.~\ref{fig3}(a) where 
$P(Q)$ for $(T',T)=(0.45,0.55)$ shows dramatically
enhanced bimodality as compared to $T'=T=0.55$.
The natural inference is that the reduction in $T'$
has resulted in an increase of the critical temperature, 
so that $T=0.55$ is now effectively deeper into the coexistence 
region. Thus, we interpret the free energy minimum in $P(Q)$ 
as an interfacial cost, now corresponding to the spatial coexistence of 
equilibrium states at temperatures $T'$ and $T$. 

We show in Fig.~\ref{fig3}(b) a similar comparison for $T=0.6$ 
where configurations $\CC_0$ at taken either from equilibrium
at $T'=0.6$, or sampled from the non-equilibrium 
$s$-ensemble at that same temperature with a biasing field towards
atypically low dynamical activity \cite{hedges2009,inactive2011}.  
In terms of inherent structure energies, these inactive 
states represent glasses with very low fictive 
temperatures, $T' \approx 0.4$ \cite{inactive2011}.
We observe again strong bimodality, in contrast to the corresponding 
equilibrium behaviour, which has nearly Gaussian fluctuations.  
The free-energy minimum in $P(Q)$ is the interfacial cost 
between low-$Q$ states (typical of the equilibrium fluid
at $T=0.6$), and high-$Q$ states (stable glassy states).
%This is illustrated in the snapshots of Fig.~\ref{fig3}(b)
%showing configurations representative of intermediate 
%values of the global overlap. The unremarkable distribution 
%found for $T'=T=0.6$ corresponds to a spatially 
%homogeneous distribution of regions with high and low overlap,
%whereas the strongly bimodal distribution observed for inactive states
%corresponds to configurations where high and low overlap 
%regions are fully demixed. 

The results in Fig.~\ref{fig3}(a,b) are significant because they 
directly evidence the existence of a free-energy cost for 
spatial coexistence between amorphous states. For 
$(T'=0.45, T=0.55)$ we obtained data for $N=150$ and $256$ and 
find that this cost increases with $N$. Additional system sizes 
are needed to infer a numerical value for $\Upsilon$, via the scaling 
$\Delta F \sim \Upsilon L^\theta$. The exponent $\theta$ is not known, 
but should obey $\theta \leq d-1$, because  
interfaces can use quenched disorder to 
optimize their geometry and reduce their cost \cite{Young_1998}. 
Physically the existence of an interfacial cost for the coexistence of 
stable glassy configurations with the equilibrium fluid, as in 
Fig.~\ref{fig4}(b), suggests that nucleation and growth is 
the appropriate mechanism to interpret the kinetic stability and 
melting dynamics of ultrastable glasses \cite{hocky2014,sepulveda2014}.  
Systematic studies along the present lines should 
help understanding kinetic stability from first-principles. 

We have presented the results of extensive 
free-energy calculations in supercooled liquids to 
assess the existence of thermodynamic singularities 
in constrained supercooled liquids, and to explore their consequences. 
Overall, our results are consistent with effective field theories 
predicting the existence of a random critical point, 
associated (below $T_c$) with phase coexistence between metastable states.
These results are also consistent with static correlations on 
length scales comparable with our largest system 
size \cite{jack2012,jack2014}. 
Whatever the status of the phase transition in the thermodynamic limit,
we find that significant static fluctuations are present
at a temperature $T \leq 0.55$, where dynamics is glassy. 
Note that if the critical point in this system is indeed 
at $T_c \approx 0.55$, a significant surface tension $\Upsilon$ 
only exists when $T \ll 0.55$. Thus, the emergence 
of activated dynamics between metastable states might 
well fall out of the dynamic range currently accessible 
to simulations. Our approach generically 
allows the determination of $\Upsilon$ in any material 
(including hard spheres), and extends to stable glassy states. 
This offers the potential for quantitative analysis of glass stability based 
on surface and bulk free energies. Overall, while 
our results do not speak directly to the mechanism of structural relaxation in 
glasses and supercooled liquids, 
they provide direct evidence that an interfacial free-energy 
barrier between metastable states is relevant both in
the `melting' of stable glasses and in glassy dynamics at equilibrum.

\acknowledgments

We thank G. Biroli, D. Coslovich and G. Tarjus for discussions.
The research leading to these results has received funding
from the European Research Council under the European Union's Seventh
Framework Programme (FP7/2007-2013) / ERC Grant agreement No 306845.
RLJ was supported by the EPSRC through grant EP/I003797/1.


\begin{thebibliography}{99}

\bibitem{Berthier_Biroli_2011}
L. Berthier and G. Biroli,  Rev. Mod Phys.  \textbf{83}, 587 (2011).

\bibitem{hansen_theory_1986}
J.-P. Hansen and I.R. {McDonald}, \emph{Theory of Simple Liquids}, 2nd ed.
  (Academic Press, London, 1986).

\bibitem{gilles}
L. Berthier and G. Tarjus, Phys. Rev. Lett. {\bf 103}, 170601 (2009).

\bibitem{rfot_review}
V. Lubchenko and P. G. Wolynes,
Annu. Rev. of Phys. Chem. {\bf 58}, 235 (2007).

\bibitem{thirumalai__1989}
T.R. Kirkpatrick, D. Thirumalai and P.G. Wolynes,  Phys. Rev. A  \textbf{40},
  1045 (1989).

\bibitem{franz_phase_1997}
S. Franz and G. Parisi,  Phys. Rev. Lett.  \textbf{79}, 2486 (1997).

\bibitem{cammarota_phase-separation_2010}
C. Cammarota, A. Cavagna, I. Giardina, G. Gradenigo, T. S. Grigera, G. Parisi
  and P. Verrocchio,  Phys. Rev. Lett.  \textbf{105}, 055703 (2010).

\bibitem{moore2012}
J. Yeo and M. A. Moore,
Phys. Rev. E {\bf 86}, 052501 (2012). 

\bibitem{berthier_overlap_2013}
L. Berthier,  Phys. Rev. E  \textbf{88}, 022313 (2013).

\bibitem{sconf}
L. Berthier and D. Coslovich,
Proc. Natl. Acad. Sci. USA {\bf 111}, 11668 (2014).

\bibitem{long_giulio}
C. Cammarota and G. Biroli,  J. Chem. Phys.  \textbf{138}, 12A547 (2013).

\bibitem{FP2013}
S. Franz and G. Parisi,  J. Stat. Mech.  p. P11912 (2013).

\bibitem{tarzia}
G. Biroli, C. Cammarota, G. Tarjus and M. Tarzia,  Phys. Rev. Lett.
  \textbf{112}, 175701 (2014).

\bibitem{kurchan93}
J. Kurchan, G. Parisi and M. A. Virasoro, J. Phys. I  \textbf{3}, 1819 (1999).

\bibitem{franz_parisi1998}
S. Franz and G. Parisi,  Physica A  \textbf{261}, 317 (1998).

\bibitem{cardenas_constrained_1999}
M. Cardenas, S. Franz and G. Parisi,  J. Chem. Phys.  \textbf{110}, 1726
  (1999).

\bibitem{Bouchaud-JCP2004}
J.-P. Bouchaud and G. Biroli,  J. Chem. Phys.  \textbf{121}, 7347 (2004).

\bibitem{chiara1}
C. Cammarota, A. Cavagna, G. Gradenigo, T. S. Grigera, and P. Verrocchio
J. Stat. Mech. L12002 (2009).

\bibitem{chiara2}
C. Cammarota, A. Cavagna, G. Gradenigo, T. S. Grigera, and P. Verrocchio,
J. Chem. Phys. {\bf 131}, 194901 (2009).

\bibitem{hedges2009}
L.~O.~Hedges, R.~L.~Jack, J.~P.~Garrahan and D.~Chandler, 
Science {\bf 323}, 1309 (2009).

\bibitem{inactive2011}
R. L. Jack, L. O. Hedges, J. P. Garrahan, and D. Chandler,
Phys. Rev. Lett. {\bf 107}, 275702 (2011). 

\bibitem{sepulveda2014}
A. Sepulveda, M. Tylinski, A. Guiseppi-Elie, R. Richert, and M. D. Ediger,
Phys. Rev. Lett. {\bf 113}, 045901 (2014).

\bibitem{hocky2014}
G. M. Hocky, L. Berthier, and D. R. Reichman,
J. Chem. Phys. {\bf 141}, 224503 (2014).

\bibitem{kob_testing_1995}
W. Kob and H. C. Andersen,  Phys. Rev. E  \textbf{51}, 4626 (1995).

\bibitem{monasson}
R. Monasson, Phys. Rev. Lett. {\bf 75}, 2847 (1995).

\bibitem{mezard99}
M. M\'ezard,  Physica A  \textbf{265}, 352 (1999).

\bibitem{garrahan2014}
J. P. Garrahan,  Phys. Rev. E  \textbf{89}, 2014 (2014).

\bibitem{parisi_liquid-glass_2014}
G. Parisi and B. Seoane,  Phys. Rev. E  \textbf{89}, 022309 (2014).

\bibitem{bomont_probing_2014}
J.-M. Bomont, G. Pastore and J.-P. Hansen, EPL \textbf{105}, 36003 (2014).

\bibitem{domb1972}
C. Domb and M. S. Green, \emph{Phase {Transitions} and {Critical} {Phenomena}.
  {Vol}. 2.}   (Academic Press, London, 1972).

\bibitem{Young_1998}
T. Nattermann, in 
\emph{Spin Glasses and Random Fields}, Ed.: A. P. Young  (World Scientific,
Singapore, 1998).

\bibitem{Fisher-PRL-1986}
D. S. Fisher,
Phys. Rev. Lett. 56, 416 (1986).

\bibitem{schwartz85}
M. Schwartz and A. Soffer,
Phys. Rev. Lett. 55, 2499 (1985).

\bibitem{science05}
L. Berthier, G. Biroli, J.-P. Bouchaud, L. Cipelletti, D. El Masri, 
D. L'Hote, F. Ladieu, and M. Pierno, Science {\bf 310}, 1797 (2005).

\bibitem{berthier_spontaneous_2007}
L. Berthier, G. Biroli, J.-P. Bouchaud, W. Kob, K. Miyazaki and D. R. Reichman,
  J. Chem. Phys.  \textbf{126}, 184503 (2007);
  J. Chem. Phys.  \textbf{126}, 184504 (2007).

\bibitem{gillesRG}
G. Tarjus, I. Balog, and M. Tissier, 
EPL {\bf 103}, 61001 (2013).

\bibitem{fytas}
N. G. Fytas and V. Martin-Mayor,
Phys. Rev. Lett. {\bf 110}, 227201 (2013). 

\bibitem{wolynesRFIM}
J. D. Stevenson, A. M. Walczak, R. W. Hall and P. G. Wolynes,
J. Chem. Phys. {\bf 129}, 194505 (2008).

\bibitem{jack2012}
R. L. Jack and L. Berthier,  Phys. Rev. E  \textbf{85}, 021120 (2012).

\bibitem{jack2014}
C. J. Fullerton and R. L. Jack, 
Phys. Rev. Lett. {\bf 112}, 255701 (2014). 

\end{thebibliography}
\end{document}